\documentclass[journal]{IEEEtran}
\ifCLASSINFOpdf
\else
   \usepackage[dvips]{graphicx}
\fi
\usepackage{url}

\usepackage{graphicx}
\usepackage[utf8]{inputenc} 
\usepackage[T1]{fontenc}
\usepackage{url}
\usepackage{bm}
\usepackage[cmex10]{amsmath} 
\usepackage{ifthen}
\usepackage{cite}
\usepackage{color}
\usepackage{graphicx}
\usepackage{amssymb}
\usepackage{amsthm}
\usepackage{amsfonts}
\usepackage{comment}
\usepackage[utf8]{inputenc} 
\usepackage[T1]{fontenc}
\usepackage{url}
\usepackage{ifthen}
\usepackage{cite}
\usepackage{color}
\usepackage{graphicx}
\usepackage{amssymb}
\usepackage{amsthm}
\usepackage{amsfonts}
\usepackage{comment}

\usepackage[skip=5pt,style=base]{caption}

\usepackage{bm}
\usepackage[cmex10]{amsmath} 

\DeclareMathOperator{\E}{\mathbb{E}}
\DeclareMathOperator{\C}{\mathbb{C}}

\DeclareMathOperator{\Tr}{Tr}
\usepackage[skip=5pt,style=base]{caption}
\begin{document}
\title{Quantification of mismatch error in randomly switching linear state-space models}


\author{Parisa Karimi, Zhizhen Zhao,  Mark Butala, and Farzad Kamalabadi
\thanks{This work was supported by the Grainger college of engineering UIUC-ZJU institute.}
\thanks{P.Karimi$^*$, Z.Zhao, and F.Kamalabadi are with the University of Illinois at Urbana-Champaign, Urbana, IL. 61801, USA (e-mail$^*$: parisa2@illinois.edu).}
\thanks{M.Butala is with the Zhejiang University, Hangzhou, Zhejiang, P.\ R.\ China.}}

\maketitle
\begin{abstract}
Switching Kalman Filters (SKF) are well known for their ability to solve the piece-wise linear dynamic system estimation problem using the standard Kalman Filter (KF). Practical SKFs are heuristic, approximate filters that are not guaranteed to have optimal performance and require more computational resources than a single-mode KF. On the other hand, applying a single-mode mismatched KF to a switching linear dynamic system (SLDS) results in erroneous estimation. This paper aims to quantify the average error an SKF can eliminate compared to a mismatched, single-mode KF in a known SLDS before collecting measurements. Mathematical derivations for the first and second moments of the estimators’ errors are provided and compared. One can use these derivations to quantify the average performance of filters beforehand and decide which filter to run in operation to have the best performance in terms of estimation error and computation complexity. We further provide simulation results that verify our mathematical derivations.
\end{abstract}

\IEEEpeerreviewmaketitle

\vspace{-.5cm}
\section{Introduction}

\IEEEPARstart{A}{} pervasive problem in virtually all branches of physics and engineering sciences, such as time-dependent tomography and imaging \cite{rem1,rem2}, geophysical data assimilation \cite{geo}, genetics \cite{gene}, and economic forecasting \cite{eco}, is the estimation of multi-dimensional state variables of a dynamical system from a collection of indirect, noisy measurements. Given the initial state distribution and a state-space model, state estimates may be recovered using Bayesian inference algorithms \cite{bayes}. The Kalman filter \cite{kalman} provides the optimal solution for the linear state-space model with additive Gaussian noise\cite{lin}.  


A computationally efficient generalization of the linear state-space model is obtained by augmenting hidden discrete random variables to the linear model to account for nonlinearities, referred to as a switching linear dynamic system (SLDS) \cite{dbn}. In this model, random switches occur in the system’s dynamic model, and Bayesian estimation may be used to estimate both the discrete modes and continuous states. Finding the exact posterior and optimal filtering in this scenario is computationally intractable \cite{SKF} since the belief state grows exponentially with time, and practical SKF formulations produce suboptimal estimates (see e.g. \cite{jump,imm,SKF}, and the references within).
This paper aims to quantify the deviation of the well-known SKF estimators from a single-mode mismatched KF analytically because the performance of the SKF may be significantly better than or close to a single-mode KF depending on the particulars of the switching distributions. This study is essential as using an SKF requires additional computation compared to a single-mode KF, and this computational burden can be significant or even intractable if the state dimension is large. The result developed in this work ensures that the SKF is used in practice only if its estimation has a considerable improvement compared to that of a mismatched KF (the metric by which an improvement is deemed significant or not is application-based). Our goal is achieved by studying the estimation errors as a function of the switching model and filter parameters, as well as the initial conditions. 

The performance of mismatched KFs in single-mode linear dynamic system has been studied previously \cite{moore,mismatched,mis2,mis3,new1}. In the case of switching dynamics, \cite{hybrid} explores the conditions under which the instantaneous mode detection is successful or not based on the statistics of the residuals of the predicted and the collected measurements. Also, Zhang et al. \cite{modebased,quant,modebased2,modebased3} study the convergence of a mode-based KF and argue the conditions under which the steady state’s bias term will converge to zero in a switching mode dynamic system. However, to the best of our knowledge, estimation of the transient evolution of the error in an SLDS prior to running the experiment and collecting measurements has not been investigated. This paper provides a quantitative measure of how effective an SKF is in an SLDS in terms of mean squared error (MSE) compared to a mismatched, single-mode KF before collecting the measurements. The result informs the decision of whether or not to use an SKF in a particular scenario and how to choose the filter with the best performance in terms of computational considerations and estimation accuracy.

The SKF algorithm’s performance in terms of MSE is a function of 1) the detection rate at each time step (and the detection algorithm), and 2) the mismatch bias whenever the algorithm detects the wrong mode. Both these parameters are functions of the switching distributions and transition probabilities. Due to space restrictions, we have assumed an estimate or an upper bound of the detection rate to be known in this work and calculate the MSE accordingly. Approximation of the detection rate as a function of the problem specification using approximate metrics \cite{binclass,bh,cher,dis} will be studied in a later publication.

The remainder of the paper is organized as follows. The common notations used throughout the paper are given in Section \ref{section:not}. The SLDS signal model and KF/SKF formulations are reviewed in Sections \ref{section:slds} and \ref{section:skf}, respectively. Section \ref{section:error} derives estimation error for mismatched KFs and SKFs in SLDS and Section~\ref{section:diss} discusses practical implementations. Simulations verify the derivations in Section~\ref{section:sim}, and conclusions are presented in Section~\ref{section:con}. 

\section{Notation}\label{section:not}
\begin{itemize}
    \item $\bm{x}\sim N(\bm{m},\bm{C})$: the random vector $\bm{x}$ has a Gaussian distribution with mean $\bm{m}$ and covariance $\bm{C}$.
\item $\E,\C,p$ refer to the expectation, covariance, and probability operators, respectively, and $\bm{I}$ is the identity matrix. 
 \end{itemize}
 \vspace{-.55cm}
\section{Signal model}\label{section:slds}
The state-space model for an SLDS may be defined as
\begin{align}
    \bm{x}_n &= \bm{A}_n \bm{x}_{n-1}+\bm{\nu}_n, \label{eqn:sp1}\\
    \bm{y}_n &= \bm{H}_n\bm{x}_n +\bm{\omega}_n \label{eqn:sp2},
\end{align}
where the subscript $n$ is the time step, $\bm{x}_n$ is the hidden state variable to be estimated, and the given model parameters are the measurement vector $\bm{y}_n$, the $z\times z$ evolution matrix $\bm{A}_n \in \{ \bm{A}^{(S_n)}\}, S_n=1,...,r$ ($S_n$ is the hidden random variable determining the mode of the system, to be detected), the $m\times z$ measurement matrix $\bm{H}_n$, and the covariance matrices $\bm{Q}_n$ and $\bm{R}_n$ where $\bm{\nu}_n \sim N(0,\bm{Q}_n)$ and $\bm{\omega}_n \sim N(0,\bm{R}_n)$ such that $\E[\bm{\nu}_n\bm{\nu}_{n'}^T]=\bm{Q}_n\,\delta(n-n')$ and $\E[\bm{\omega}_n\bm{\omega}_{n'}^T]=\bm{R}_n\,\delta(n-n')$, and $\E[\bm{\omega}_n\bm{\nu}_{n'}^T]=\bm{0}$ ($m$ is the number of measurements, $z$ is the state dimension, and $r$ is the number of modes the system may switch between). 
\section{Kalman filter/Switching Kalman filter}\label{section:skf}
 The KF is an optimal estimator for linear dynamic systems. Let $\bm{y}_1^n\equiv [\bm{y}_1,\bm{y}_2,...,\bm{y}_n]$, $\bm{x}_0\sim N(\bm{x}_{0|0},\bm{P}_{0|0})$, $\bm{x}_{n|n}=\E[\bm{x}_n|\bm{y}_1^n], and \bm{P}_{n|n}=\C(\bm{x}_n|\bm{y}_1^n)$. The ``$Filter$'' operator is defined as
\begin{multline}
    (\bm{x}_{n|n},\bm{P}_{n|n}) = Filter(\bm{A}_n,\bm{H}_n,\bm{x}_{n-1|n-1},\bm{P}_{n-1|n-1}, \\ \bm{Q}_n,\bm{R}_n,\bm{y}_1^n), \label{eqn:filter}
\end{multline}
which involves the repeated application of a time update step
\begin{align}
\bm{x}_{n|n-1} &= \bm{A}_n \bm{x}_{n-1|n-1}, \label{eqn: Kalman1}\\
\bm{P}_{n|n-1}&= \bm{A}_n \bm{P}_{n-1|n-1}\bm{A}_n^T + \bm{Q}_n;
\end{align}
and a measurement update step
\begin{alignat}{3}
\bm{\epsilon}_n &= \bm{y}_n - \bm{H}_n \bm{x}_{n|n-1},\; & \bm{B}_n &= \bm{H}_n \bm{P}_{n|n-1} \bm{H}_n^T + \bm{R}_n,\\
\bm{K}_n &= \bm{P}_{n|n-1} \bm{H}_n^T \bm{B}_n^{-1},\; &
\bm{x}_{n|n} &= \bm{x}_{n|n-1} + \bm{K}_n \bm{\epsilon}_n,\\
&& \bm{P}_{n|n} &= (\bm{I} - \bm{K}_n \bm{H}_n) \bm{P}_{n|n-1}. \label{eqn:Kalmanend} 
\end{alignat}
  The application of a single-mode KF to the general SLDS of \eqref{eqn:sp1}-\eqref{eqn:sp2} results in erroneous estimates. The well known SKF formulation detects the switching mode and its corresponding model parameters ($\bm{A}_n,\bm{Q}_n$) at each time step, and estimates the state variables accordingly. Upon perfect detection of the modes, one could obtain optimal estimates of the state variable $\bm{x}_n$ in terms of both MAP and MSE metrics. Due to the exponentially explosion of the posterior in optimal SKFs\cite{SKF}, several approximate SKF algorithms have been proposed (e.g.\ \cite{SKF,imm}). Due to space restrictions, we assumed an estimate of the detection rate to be known and the details of the approximate SKF algorithms are not presented here. 

\section{Derivation of the mean squared error}\label{section:error}
To quantify the performance of each filter, we first study the estimation error imposed by applying a mismatched model to a single-mode linear dynamic model in Section~\ref{section:mismatch}. Then, in Sections~\ref{section:single} and \ref{section:switch}, the effect of applying a single-mode mismatched KF and SKF to an SLDS is quantified using prior and transition probabilities of the SLDS.
\subsection{Mismatched Kalman filter error}\label{section:mismatch}
Instead of the state-space equations \ref{eqn:sp1}-\ref{eqn:sp2} with the correct dynamic evolution model $(\bm{A}_n, \bm{Q}_n)=(\bm{A},\bm{Q})$ at time $n$, consider a mismatched model using $(\bm{A}_n^d, \bm{Q}_n^d)=(\bm{A}^d, \bm{Q}^d)$ (superscript $d$ refers to the mismatched model). It is well known that KF estimates are unbiased and optimal, but this is true only when the correct model is used. In order to determine how far the estimates are from the correct model estimates, we study the error term $\bm{e}_n=\bm{x}_n-\bm{x}_{n|n}$, where $\bm{x}_n$ is the ground truth state variable, $\bm{x}_{n|n}=\bm{x}_{n|n}^d$ is associated with the mismatched model, and $\bm{K}_n^d=\bm{K}^d$ is the Kalman gain at time $n$ obtained based on equations \eqref{eqn: Kalman1}-\eqref{eqn:Kalmanend} using the  mismatched dynamic model $(\bm{A}^d, \bm{Q}^d)$.
\begin{align}
\bm{x}_n&=\bm{A}\bm{x}_{n-1}+\bm{\nu}_n ,\\
\bm{x}_{n|n}^d&=\bm{A}^d\bm{x}_{n-1|n-1}+\bm{K}^d(\bm{y}_n-\bm{H}_n\bm{x}_{n|n-1}^d) \nonumber\\
&=\bm{A}^d\bm{x}_{n-1|n-1}+\bm{K}^d\bigl[\bm{H}_n(\bm{A}\bm{x}_{n-1} + \bm{\nu}_n) + \bm{\omega}_n \nonumber\\
& \quad -\bm{H}_n\bm{A}^d\bm{x}_{n-1|n-1}\bigr],\\
\bm{e}_n&=\bm{A}\bm{x}_{n-1}+\bm{\nu}_n - \bm{A}^d\bm{x}_{n-1|n-1}-\bm{K}^d\bigl[\bm{H}_n(\bm{A}\bm{x}_{n-1}\nonumber\\
 & \quad +\bm{\nu}_n)+\bm{\omega}_n-\bm{H}_n\bm{A}^d\bm{x}_{n-1|n-1}\bigr]\nonumber\\
&=[\bm{A}-\bm{A}^d+\bm{K}^d\bm{H}_n\bm{A}^d-\bm{K}^d\bm{H}_n\bm{A}]\bm{x}_{n-1}\nonumber\\
&\quad +[\bm{A}^d -\bm{K}^d\bm{H}_n\bm{A}^d]\bm{e}_{n-1}+\bm{B}_n^d\bm{\nu}_n-\bm{K}^d\bm{\omega}_n \nonumber \\
&=\bigl[ \bm{B}_n^d(\bm{A} - \bm{A}^d)\bigr]\bm{x}_{n-1} 
 + \bm{B}_n^d(\bm{A}^d\bm{e}_{n-1}+\bm{\nu}_n)-\bm{K}^d\bm{\omega}_n. \label{eqn:bias}
\end{align}
where $\bm{B}_n^d = \bm{I}-\bm{K}^d\bm{H}_n$. This defines a new state space model where the noise terms are white Gaussian (but note that the measurement noise $\bm{\omega}_n$ and state evolution noise $\bm{B}_n^d\bm{\nu}_n-\bm{K}^d\bm{\omega}_n$ are dependent) and the input is $\bigl[ \bm{B}_n^d(\bm{A} - \bm{A}^d)\bigr]\bm{x}_{n-1}$.

This error term may be studied in terms of its mean and covariance. The mean is given by
\begin{equation}
    \E[\bm{e}_n] =\bm{B}_n^d(\bm{A}-\bm{A}^d)\E[\bm{x}_{n-1}] 
    + \bm{B}_n^d\bm{A}^d\E[\bm{e}_{n-1}].
\end{equation}
To calculate $\C(\bm{e}_n)$, we need to first calculate the following covariance terms: 
\begin{align}
    \C(\bm{e}_0)&=\C(\bm{x}_0-\bm{x}_{0|0})=\bm{P}_0,\quad \C(\bm{x}_0) =\bm{P}_0, \nonumber \\
    \C(\bm{x}_n)&=\C(\bm{A}\bm{x}_{n-1}+\bm{\nu}_n)=\bm{A}\C(\bm{x}_{n-1})(\bm{A})^T+\bm{Q}_n, \nonumber \\
    \C(\bm{e}_n,\,\bm{x}_n)&=\C(\bm{x}_n-\bm{x}_{n|n}^d,\,\bm{x}_n)=\C(\bm{x}_n)-\C(\bm{x}_{n|n}^d,\,\bm{x}_n). \nonumber 
\end{align}
We denote $\C(\bm{x}_{n|n}^d,\,\bm{x}_n)$ by $\bm{u}_n$ and obtain the following recursion,
\begin{equation*}
\begin{split}
\bm{u}_n&=\C(\bm{A}^d\bm{x}_{n-1|n-1}^d+\bm{K}^d(\bm{y}_n-\bm{H}_n\bm{x}_{n|n-1}^d),\,\bm{A}\bm{x}_{n-1}+\bm{\nu}_n)\nonumber\\
    &=\C(\bm{A}^d\bm{x}_{n-1|n-1}^d+\bm{K}^d(\bm{H}_n\bm{A}\bm{x}_{n-1}+\bm{H}_n\bm{\nu}_n+ \bm{\omega}_n \nonumber\\
     &\quad   -\bm{H}_n\bm{A}^d\bm{x}_{n-1|n-1}^d),\,\bm{A}\bm{x}_{n-1}+\nu_n)\nonumber\\
    &=\bm{B}_n^d\bm{A}^d\bm{u}_{n-1}\bm{A}^T +\bm{K}^d\bm{H}_n\bm{A}\C(\bm{x}_{n-1})\bm{A}^T +\bm{K}^d\bm{H}_n\bm{Q}_n,\\
    \bm{K}^d &= \bigl[\bm{A}^d\bm{P}_{n-1|n-1}(\bm{A}^d)^T+\bm{Q}^d\bigr]\\
    & \quad \times \bm{H}_n^T\bigl(\bm{H}_n\bigl[\bm{A}^d\bm{P}_{n-1|n-1}(\bm{A}^d)^T+\bm{Q}^d\bigr]\bm{H}_n^T+\bm{R}_n\bigr)^{-1},
\end{split}
\end{equation*}
with $\bm{u}_0=\C(\bm{x}_{0|0},\,\bm{x}_{0|0}+\bm{\nu}_0)=\bm{0}$. Therefore, using \eqref{eqn:bias} and calculating the covariance, we have
\begin{multline}
    \C(\bm{e}_n)=\bm{K}^d\bm{R}_n(\bm{K}^d)^T+\bm{J}_n\C(\bm{x}_{n-1})\bm{J}_n^T\\
    +\bm{B}_n\bm{Q}_n\bm{B}_n^T +\bm{B}_n\bm{A}^d \C(\bm{e}_{n-1})(\bm{A}^d)^T\bm{B}_n^T \\
    +\bm{J}_n\C(\bm{x}_{n-1},\,e_{n-1})\bm{B}_n^T + \bm{B}_n\C(\bm{x}_{n-1},\,e_{n-1}) \bm{J}_n^T.
\end{multline}
where $\bm{J}_n = \bm{A}-\bm{A}^d+\bm{K}^d\bm{H}_n\bm{A}^d-\bm{K}^d\bm{H}_n\bm{A}$. All the above variables can be calculated recursively.

\subsection{Single-mode Kalman filter error in an SLDS}\label{section:single}
In this section, an arbitrary single-mode KF is applied to an SLDS, and the MSE is calculated. Let $l_n$ refer to a trajectory from the set of all possible $r^n$ trajectories of discrete modes that may occur, where $r$ is the number of possible modes to occur at each time step and $n$ is the time step. Also, let $\bm{e}_n^{l_n}$ be the conditional error of the KF with trajectory $l_n$, so its mean and covariance can be calculated based on Section~\ref{section:mismatch} recursively given the trajectory. Assuming $l_n=[l_{n-1},i]$ s.t. $i\in \{1,...,r\}$ and $n>1$, the error at each time is given by: 
\begin{align}
    \bm{e}_n^{l_n}&=\bigl[(\bm{I}-\bm{K}_n\bm{H}_n)(\bm{A}^i-\bm{A})\bm{x}_{n-1}^{l_{n-1}}-\bm{K}_n\bm{\omega}_n\nonumber\\
    &\quad+(\bm{I}-\bm{K}_n\bm{H}_n)(\bm{A}\bm{e}_{n-1}^{l_{n-1}} + \bm{\nu}_n)\bigr], \label{eqn:singleei} \\
\bm{e}_n&=\sum_{l_n}\delta_{l_n} \bm{e}_n^{l_n}, \label{eqn:single}
\end{align}
where $\bm{x}_{n-1}^{l_{n-1}}$ and $\bm{e}_{n-1}^{l_{n-1}}$ refer to the ground truth state variable and error for trajectory $l_{n-1}$, $\bm{K}_n$ is the Kalman gain for the single-mode KF at time $n$, and $\delta_{l_n}$ equals one when trajectory $l_n$ occurs and is zero otherwise. The expectation of the error at $n$ over all possible trajectories is,
\begin{align}
\E[\bm{e}_n]&=\sum_{l_n} \pi_{l_n}\E[\bm{e}_n^{l_n}], \label{eqn:singlemean}
\end{align}
 where $\pi_{l_n}$ is the probability of trajectory $l_n$ and can be calculated based on the given SLDS transition probabilities and priors. Similarly, we compute $\E[\bm{e}_n\bm{e}_n^T]=\sum_{l_n}\pi_{l_n}\E[\bm{e}_n^{l_n}{(\bm{e}_n^{l_n})}^T]$ and the covariance $ \C(\bm{e}_n) =\E[\bm{e}_n\bm{e}_n^T]-\E[\bm{e}_n]\E[\bm{e}_n]^T$.
This formulation can be used to calculate the performance of an arbitrary single-mode KF in an SLDS. 
\subsection{Switching Kalman filter error in SLDS}\label{section:switch}
We now calculate the MSE when a SKF algorithm is applied to a known SLDS. Let $l_n$ and $q_n$ refer to the trajectory that occurs (the true trajectory) and that is detected using the SKF algorithm in an SLDS, respectively, each taking values in the set of all possible $r^n$ trajectories of length $n$ such that $l_n=[l_{n-1},i]$ , $q_n=[q_{n-1},j]$, where $i,j\in \{1,...,r\}$. Also, let $\bm{e}_n^{(l_n;q_n)}$ be the conditional error based on these trajectories, which can be calculated recursively based on results from Section~\ref{section:mismatch} given the trajectories $l_n$ and $q_n$: 
\begin{align}
    \bm{e}_n^{(l_n;q_n)}&=([\bm{A}^{i}-\bm{A}^j+\bm{K}_n^j\bm{H}_n\bm{A}^j-\bm{K}_n^j\bm{H}_n\bm{A}^{i}]\bm{x}_{n-1}^{l_{n-1}}\nonumber\\
    &\quad+(\bm{I}-\bm{K}_n^j\bm{H}_n)(\bm{A}^j\bm{e}_{n-1}^{(l_{n-1};q_{n-1})}+\bm{\nu}_n) -\bm{K}_n^j\bm{\omega}_n),
    \label{eq:switch_e_rec}
\end{align}
where $\bm{x}_{n-1}^{l_{n-1}}$ and $\bm{e}_{n-1}^{(l_{n-1};q_{n-1})}$ refer to the ground truth state and error when trajectory $l_{n-1}$ occurs and trajectory $q_{n-1}$ is detected, $\bm{K}_n^j$ is the KF gain assuming mode $j$ is detected at time $n$. The error then may be written as
\begin{align}
    \bm{e}_n =& \sum_{l_n}\sum_{q_n}\delta_{l_n;q_n}\bm{e}_n^{(l_n;q_n)} , \label{eqn:switche}
\end{align}
where $\delta_{l_n;q_n}$ equals one when $l_n$ occurs and $q_n$ is detected, and 0 otherwise. The mean of this random process is calculated as
\begin{align}
    \E[\bm{e}_n]&=\sum_{l_n}\sum_{q_n}\pi_{l_n,q_n}\E[\bm{e}_n^{(l_n;q_n)}],
\end{align}
 where $\pi_{l_n,q_n}$ is the probability that trajectory $l_n$ occurs and trajectory $q_n$ is detected, which may also be calculated recursively. Similarly, we compute $\E[\bm{e}_n\bm{e}_n^T]=\sum_{l_n}\sum_{q_n}\pi_{l_n,q_n}\E[\bm{e}_n^{(l_n;q_n)}{(\bm{e}_n^{(l_n;q_n)})}^T]$ and the covariance $ \C(\bm{e}_n) =\E[\bm{e}_n\bm{e}_n^T]-\E[\bm{e}_n]\E[\bm{e}_n]^T$.
\vspace{-.35cm}
\section{Discussion}\label{section:diss}
\vspace{-.2cm}
 Some challenges in calculating the derived statistics are discussed below. \\
 1) Applying the formulation to multi-modal cases enables making the optimal decision on which modes to keep in a SKF framework, but at huge computational cost due to having $r^n$ trajectories at time $n$ ($r$ = number of modes). A suboptimal, feasible solution to the problem calculates the marginal transition probability between each pair of modes and applies the formulation to each pair. In this framework, the collection of switching dynamic systems is represented using a graph network where each node is a dynamic system mode, as shown in \ref{fig:rmodal}. For each pair of nodes: if using an SKF for the pair does not provide a significant improvement over single-mode KF using the proposed formulation, merge them and if not, keep both. A multi-modal problem is divided into multiple bi-modal problems, as a result. \\
2) The recursive calculation of (\ref{eqn:singleei}) and (\ref{eq:switch_e_rec}) for all possible trajectories is practically infeasible as $n$ becomes large. We propose a solution to solve this problem under two scenarios: \\
\textbf{(i)} If the transition probabilities are equal, the bias at time $n$ based on (\ref{eqn:switche}) for a bi-modal system can be derived recursively using the same notation as in Section \ref{section:switch}:
\begin{figure}
  \centering
    \includegraphics[width=0.25\textwidth]{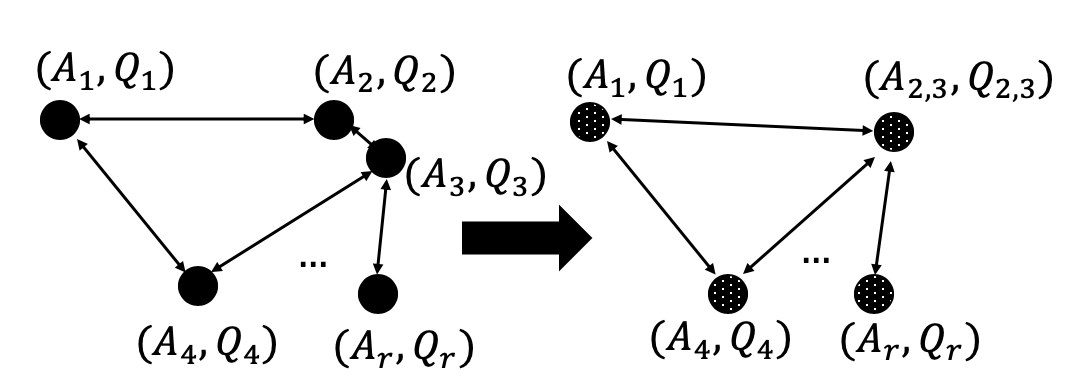}
    \caption{ SLDS model with r modes is reduced to SLDS model with (r-1) modes using the proposed metrics. }
    \label{fig:rmodal}
        \vspace{-0.4cm}
\end{figure}
\begin{align}
    \bm{e}_n &= \sum_{[l_{n-1},i]}\sum_{[q_{n-1},j]}\delta_{[l_{n-1},i];[q_{n-1},j]}\bm{e}_n^{([l_{n-1},i];[q_{n-1},j])} \nonumber\\
    &=\sum_{[l_{n-1},i]}\sum_{q_{n-1}}\delta_{TD}[n] \delta_{[l_{n-1},i];q_{n-1}}\nonumber\\
     &\quad \times L_{TD,i}(\bm{e}_{n-1}^{([l_{n-1};q_{n-1}]},\bm{\nu}_{n-1},\bm{\omega}_{n-1}) \nonumber\\
    &\quad + \sum_{[l_{n-1},i]}\sum_{q_{n-1}} \delta_{FD}[n] \delta_{[l_{n-1},i];q_{n-1}}\nonumber\\
    &\quad \times L_{FD,i}(\bm{e}_{n-1}^{([l_{n-1};q_{n-1}]},\bm{x}_{n-1}^{l_{n-1}},\bm{\nu}_{n-1},\bm{\omega}_{n-1})\nonumber\\
    &=\sum_{[l_{n-1},i]}\sum_{q_{n-1}} \delta_{i|l_{n-1}}\delta_{l_{n-1};q_{n-1}}\nonumber\\
    &\quad \times \delta_{TD}[n]  L_{TD,i}(\bm{e}_{n-1}^{(l_{n-1};q_{n-1}},\bm{\nu}_{n-1},\bm{\omega}_{n-1}) \nonumber\\
    &\quad + \sum_{[l_{n-1},i]}\sum_{q_{n-1}}  \delta_{i|l_{n-1}}\delta_{l_{n-1};q_{n-1}}\nonumber\\
    &\quad \times \delta_{FD}[n] L_{FD,i}(\bm{e}_{n-1}^{l_{n-1};q_{n-1}},\bm{x}_{n-1}^{l_{n-1}},\bm{\nu}_{n-1},\bm{\omega}_{n-1})\nonumber\\
    &=\delta \sum_{i} \delta_{TD}[n]  L_{TD,i}(\sum_{l_{n-1},q_{n-1}}\delta_{l_{n-1};q_{n-1}}\nonumber\\
    &\quad \times \bm{e}_{n-1}^{[l_{n-1};q_{n-1}]},\bm{\nu}_{n-1},\omega_{n-1}) \nonumber\\
    &\quad +\delta\sum_{i} \delta_{FD}[n] L_{FD,i}(\sum_{l_{n-1},q_{n-1}}  \delta_{l_{n-1};q_{n-1}}\nonumber\\
    &\quad \times\bm{e}_{n-1}^{l_{n-1};q_{n-1}},\sum_{l_{n-1}}\delta_{l_{n-1}}\bm{x}_{n-1}^{l_{n-1}},\bm{\nu}_{n-1},\bm{\omega}_{n-1}) \nonumber\\
    &= \delta\times\delta_{TD}[n] \sum_{i}  L_{TD,i}(\bm{e}_{n-1},\nu_{n-1},\bm{\omega}_{n-1}) \nonumber\\
    &\quad +\delta\times\delta_{FD}[n]\sum_{i}  L_{FD,i}(\bm{e}_{n-1},\bm{x}_{n-1},\bm{\nu}_{n-1},\bm{\omega}_{n-1})
\end{align}
where we assumed $\delta_{TD}[n]$ to be the Kronecker delta when the correct mode is detected at time $n$ and $\delta_{FD}[n] = 1 - \delta_{TD}[n]$ (in this study, the true detection rate is approximated by a constant rate for both modes at each time step in order to make the computations practically feasible), $\delta_{[l_{n-1},i];q_{n-1}}=\delta_{i|l_{n-1}}\delta_{l_{n-1};q_{n-1}}$ due to the Markov property of transitions, $\delta_{i|l_{n-1}}$ is the Dirac delta when mode $i$ occurs at time $n$ given that the trajectory $l_{n-1}$ occurred at time $n-1$, and $L_{TD,i}$ and $L_{FD,i}$ refer to the linear functions based on (\ref{eqn:bias}) when true detection and false detection occurs, respectively. The fact that modes $i=1,2$ are independent (so the covariance between the terms of the summation over the current mode $i$ is zero), $\delta_{i|l_{n-1}}=\delta \forall i$ due to the transitions being equi-probable, as well as the properties of covariance for a linear combination of variables, allowed us to conclude that knowledge of the mean and covariance of $\bm{x}_{n-1}$ and $\bm{e}_{n-1}$ and the noise statistics is sufficient to recursively calculate the MSE. In this case, propagating all trajectories is not required to calculate the statistics at each time step. The same reasoning also applies to the formulation of a single-mode KF in an SLDS, as a special case of this general scenario.\\
\textbf{(ii)} When the transition probabilities are not equal, the probability of each trajectory is a deterministic function of the transition matrix and the initial probabilities. Therefore, by keeping the $K$ trajectories with the largest probabilities that sum to $P_c$ and ignoring the rest of them, the calculated MSE is ensured to be in an $\epsilon$-neighborhood of the correct MSE ($\epsilon$ is a function of $P_c$ and the dynamic system's parameters). The number of trajectories required to keep increases as time increases and as the transition probabilities tend to equal ($0.5$).

Mode detection in an SKF has its largest error when all transition probabilities are equal (0.5 in bimodal system), since the uncertainty is maximized in this case. Therefore, studying the performance of an SKF using equal transition probabilities between modes is a computationally feasible and robust metric for the purpose of comparison of the performance of an SKF with a single-mode Kalman filter.

 \vspace{-.35cm}
\section{Simulations}\label{section:sim}
A bi-modal 4D state-space model is simulated via Monte Carlo (MC) simulations, and the error statistics calculated using analytic derivations and MC simulations are compared to verify the proposed formulation. An approximation of the detection rate is assumed to be given in these simulations and the SKF gain at time $n$ when mode $i$ occurs is approximated by the gain of the KF with mode $i$ at time $n$. Let the state-space equations be as presented in (1)-(2) such that $\bm{A}^{(1)}=0.9\bm{I}_{4\times 4}$, ${\bm{A}}^{(2)}=0.46\bm{I}_{4\times 4}$, $\bm{Q}^{(1)} =\bm{Q}^{(2)} =0.01\bm{I}_{4\times 4}$, $\bm{R}=0.01\bm{I}_{4\times 4}$, $\bm{x}_0=[1,1,1,1]^T$, and $\bm{P}_0 = \bm{I}_{4\times 4}$, and $\bm{H}=\bm{I}_{4\times 4}$ with the transition matrix $\bm{Z}=[0.5,0.5;0.5,0.5]$ and the prior $\pi=[0.5,0.5]$.

 The single-mode KFs using models (1) and (2), as well as the ``average KF'' $\bm{A}_n = \pi_1 \bm{A}^{(1)} + \pi_2 \bm{A}^{(2)}$, where $\pi_i$ is the probability of mode $i$ at each time $n$ (calculated based on the prior and transition probabilities of the discrete mode), are used for estimation. Intuitively, if the switching distributions are close to each other (e.g. in the KL divergence sense), the average filter’s estimates are close to the optimal solution. Alternatively, if the distributions are far from each other (in the KL divergence sense), the average KF’s estimates are poor. 

\begin{figure}
  \centering
    \includegraphics[width=0.3\textwidth]{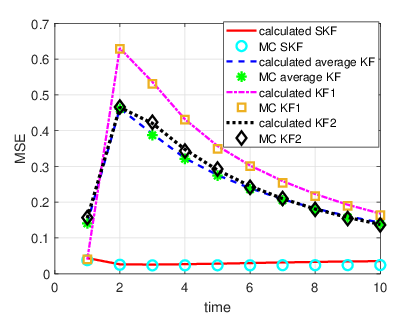}
    \caption{ MC obtained MSEs verify the derivations.}
    \label{fig:ex}
    \vspace{-0.4cm}
\end{figure}

Using 20k MC simulations, $\E[\bm{e}_n]$ and $\C[\bm{e}_n]$ are calculated analytically, and the MSE is calculated based on $\text{MSE}=E[e_n]^TE[e_n]+\Tr [\C(e_n)]$ ($\Tr$ refers to the trace of the matrix) for the SKF and mismatched single-mode KFs using the proposed formulation. Fig.~\ref{fig:ex} show the consistency between the analytic and MC calculated MSEs. In this case, there is a visually significant difference between the performance of the SKF compared to the single-mode KFs (the level of significance is application dependent). Simulations for higher state dimensions also verified the MSE derivations.

\vspace{-.3cm}
\section{Conclusion and future work}\label{section:con}
The MSE performance of the SKF was compared to an arbitrary single-mode mismatched KF analytically using a recursive formulation. This formulation may be used to decide which filter to run operationally for a specific SLDS. This work is a step towards automating the filter decision process for a specific SLDS scenario by evaluating the accuracy versus computation trade-off.
\bibliographystyle{IEEEtran}
\bibliography{SPL}{}










\end{document}